\documentclass[twoside,12pt]{article}
\usepackage{epsf,epsfig,amssymb,amsmath,graphicx,float}
\usepackage{color}

\begin{document}
\renewcommand{\thefootnote}{\fnsymbol{footnote}}

\begin{titlepage}

\begin{center}

\vspace{1cm}

{\Large {\bf Relic density of asymmetric dark matter and constraints on the
    parameter spaces}}

\vspace{1cm}

{\bf Qiu Sujuan, Hoernisa Iminniyaz}\footnote{Corresponding 
author, wrns@xju.edu.cn}

\vskip 0.15in
{\it
{School of Physics Science and Technology, Xinjiang University, \\
Urumqi 830017, China} \\
}

\abstract{Based on the cosmological observations, the baryon and dark matter
  energy densities are similar. There may exist an asymmetry between the dark
  matter particles and their
  corresponding anti--particles. We review the relic density of asymmetric dark
  matter particle and anti--particle in the standard cosmological scenario and  quintessence model with kination phase.
  Then we use the Planck data to find constraints on the annihilation cross
  section and the mass of the asymmetric dark matter in those models.
 }
\end{center}
\end{titlepage}
\setcounter{footnote}{0}

\section{Introduction}
Astronomical and cosmological observations have demonstrated the existence of
non--luminous matter \cite{Planck:2018vyg}. Measurements of the
anisotropy of cosmic microwave
background radiation have enabled accurate determinations of the total amount
of dark matter \cite{Planck:2018vyg}.  The energy density of dark matter and
baryonic matter (visible matter, which participates in electromagnetic,
strong and weak interactions) constituting the present--day
universe have been determined from the observations by Planck satellite data
\cite{Planck:2018vyg},
\begin{equation}\label{eq:omega}
\Omega_{\rm DM} h^2 = 0.120\pm 0.001\,,\,\,\,             
\Omega_{\rm B} h^2 =0.0224 \pm 0.0001 \,,
\end{equation}
where $\Omega_{\rm DM}$ represents the relic density of dark matter in units
of the critical density, and
$h$ is the Hubble constant, $h=0.7100 \pm 0.025$, measured
in units of 100 kilometers per second per megaparsec, $\Omega_{\rm B}$
is the relic density of baryonic matter.

Although we have an accurate measurement of the dark matter relic density, the
physical properties of dark matter in the universe are still unknown. In order
to detect the dark matter particles, many methods are used and ongoing, such
as direct detection, indirect detection and particle collision experiments. In
theoretically, candidate particles for dark matter are proposed beyond the
Standard Model of particle physics. Weakly interacting massive particles
(WIMPs) are the most interesting candidates for dark matter, and neutralino is
the best motivated one which is appeared in supersymmetric models. Neutralinos
are the Majorana particles which are their own anti--particles. Untill now, we
have no evidence that the dark matter particle should be Majorana
particle. There is another option that the dark matter particles can be
asymmetric. On the other hand, from Eq.(\ref{eq:omega}), we know
$\Omega_{\rm DM} \approx 5 \Omega_B$. It motivates us to consider whether
there is connection between the dark matter and the baryonic matter. There maybe
similar origin which can explain the postulated asymmetry for the dark matter and baryon asymmetry.

Asymmetric dark matter has been widely studied in the literatures
\cite{adm-models,frandsen,Petraki:2014uza}. In
many current models, asymmetry arises in one sector, either in the Standard Model
sector or in the dark matter sector, and is subsequently transmitted via contact
to the other sector. This situation often leads to similar baryon and dark
matter number densities, with $n_{DM} \sim n_b$. The symmetric component of
dark matter number density is typically eliminated in the hidden region,
and the relic density of dark matter is determined mainly by the asymmetric
components.
Asymmetric dark matter abundance is obtained by solving the Boltzmann
equations for particle and anti--particle which describes the time evolution of
particle and anti--particle densities in the expanding universe
\cite{Iminniyaz:2013cla}. This
kind of computations have been done for the standard cosmology and
non--standard cosmological models in refs.\cite{GSV,Iminniyaz:2011yp,Iminniyaz:2018das}. In the context of asymmetric,
weakly interacting massive particles, the relic density of asymmetric dark matter depends on its thermal annihilation cross--section, mass, and the existed asymmetry.

In our work, we review the calculation of the relic density of asymmetric WIMP
dark matter in the standard cosmological scenario and quintessence model with
kination phase. The Hubble parameter is
modified in quintessence model with kination phase and the relic density of dark matter is
affected by the modification. The detailed analysis has been done in
refs.\cite{Salati:2002md,Schelke:2006eg,Catena:2009tm}. We use the observed abundance of dark matter
to derive the constraints on the parameter spaces including the mass and cross
sections for asymmetric WIMP dark matter in the standard and quintessence models. We plot the contour of the mass and cross section when the dark matter relic density satisfies the observational value. We find when the cross section is small, the relic density is insensitive for the varying mass. When the mass is increased to the maximum value, the final relic abundance of asymmetric dark matter is not much affected for the varying cross section in both standard and quintessence models. Because of the increased Hubble expansion rate in quintessence model, one needs larger annihilation cross section than the standard one in order the relic density falls in the observation range.

The structure of our paper is as follows. In section 2, we briefly review the
Boltzmann equations for asymmetric dark matter in the standard cosmological
model. In section 3, we discuss the Boltzmann equations with an altered Hubble
parameter in the quintessence model for particles and anti--particle. In section 4, we
derive the constraints on the parameter spaces including the cross section and the
mass. The conclusion and summaries are in the
last section.

\section{Boltzmann equations for asymmetric dark matter in the standard
  cosmological model }

In this section, we briefly review the Boltzmann equations for asymmetric
dark matter in the standard cosmological scenario. We assume that the
asymmetric dark matter particles and anti--particles were in thermal
equilibrium in the early universe and they were decoupled when they are in
non--relativistic case. The number densities of asymmetric dark matter in the
expanding universe evolved over time are described by the Boltzmann equations.
By solving the Boltzmann equations, we obtain
the relic density of asymmetric dark matter for particles and anti--particles.
Here we consider that only particles $\chi$ and $\bar\chi$ can be
annihilated into the Standard Model particles, $\chi \chi$ and
$\bar\chi \bar\chi$ can not, then the general form of the Boltzmann equations
for the particle $\chi$ and anti--particle $\bar\chi$ is
\begin{eqnarray}\label{eq:boltzmann_n}
\frac{d n^{\pm}}{dt} + 3 H n^{\pm} &=&
- {\langle \sigma v\rangle} (n^{+} n^{-} -
n^+_{\rm eq} n^-_{\rm eq})\,,
\end{eqnarray}
where $n^+$ is for particle and $n^-$ for anti--particle, $\langle \sigma v \rangle$ is the thermal average of the annihilation
cross section multiplied by the relative velocity of the two annihilating
particles. During the radiation--dominated period, the Hubble parameter
$H = \pi T^2/M_{\rm Pl} \, \sqrt{g_*/90}$,
   where $M_{\rm Pl} =2.4 \times 10^{18}$ GeV is the Planck mass and $g_*$ the
effective number of relativistic degrees of freedom. The number densities $n_{eq}^+$ and
$n_{eq}^-$ of asymmetric dark matter in thermal equilibrium at
non--relativistic limits can be written as
\begin{equation}\label{eq:neq}
n^{\pm}_{\rm eq}= g\, {\left[m T/(2 \pi) \right]}^{3/2}{\rm e}^{-m/T {\pm \,\xi}}\,,
\end{equation}
where $g$ represents the number of internal degrees of
freedom and m is the mass of the dark matter particle,
$\xi = \mu/ T$, here $\mu$ is the chemical potential
\cite{GSV,Iminniyaz:2011yp,standard-cos}. In the standard
cosmological scenario, for $T \gg m$, asymmetric dark matter particles
and anti--particles are in the thermal equilibrium state.
In later time when $T \ll  m$, the annihilation rate
$\Gamma = n_{\chi} \langle \sigma v \rangle$ drops below the expansion rate
$H$, therefore, the
particles can not be effectively annihilated and the co--moving number
densities become constant. The temperature at which particles
decoupled from the hot bath is called the freeze out temperature
$T_{F}$.

The Boltzmann Eq.(\ref{eq:boltzmann_n}) is simplified by
introducing the dimensionless quantities $Y^+ = n^+/s$, $ Y^- =n^-/s$
and variable $x = m/T$, where the entropy density
$s = 2 \pi^2 g_*/45\,T^3$. Suppose that when the universe is in a state of
adiabatic expansion, the entropy of each co--moving volume is conserved.
The Boltzmann equation (\ref{eq:boltzmann_n}) can be rewritten as
\begin{eqnarray} \label{eq:boltzmann_Y}
\frac{d Y^{\pm}}{dx}
    = - \frac{\lambda\langle \sigma v \rangle }{x}\,
     (Y^+~Y^{-} - Y^+_{\rm eq}~Y^-_{\rm eq}) \,,
\end{eqnarray}
where $\lambda = 1.32\,m M_{\rm Pl}\, \sqrt{g_*}$. The detailed derivation of Eq.(\ref{eq:boltzmann_Y}) is found in ref.\cite{adm-models}. Subtracting
equation for $Y^-$ from the equation for $Y^+$,
we obtain $dY^+/dx - dY^- / dx = 0$. It yields $Y^+ - Y^- = \eta$. Here $\eta$
is considered as a primordial asymmetry between the particle and anti--particle.
The dark matter asymmetry is expressed in units of the baryon asymmetry as
$\eta =  \epsilon {\eta_{B}}$, where
$\eta_{B} = (0.880 \pm 0.021) \times 10^{-10}$ \cite{WMAP:2010qai}. The Boltzmann equation can
be rewritten as follows,
\begin{equation} \label{eq:Yeta}
\frac{d Y^{\pm}}{dx} =
     - \frac{\lambda \langle \sigma v \rangle}{x^2}~
     \left[(Y^{\pm})^2 \mp \eta Y^{\pm}  - Y_{\rm eq}^2\right]\,,
\end{equation}
where
$Y^2_{\rm eq}= Y^+_{\rm eq} Y^-_{\rm eq}=(0.145g_{\chi}/g_*)^2\,
x^3e^{-2x}$.  These equations can be solved through numerical and analytical
methods.

Generally, the annihilation cross--section of WIMPs is expanded in the
relative velocity $v$ of the annihilating WIMP particles \cite{standard-cos} as
\begin{eqnarray} \label{eq:langle}
\langle \sigma v \rangle =a + b \langle v^2 \rangle + O (\langle v^4 \rangle)
= a + 6 b x^{-1} + O (x^{-2})\,.
\end{eqnarray}
Here $a$ is $s–$wave contribution to $\sigma v$ while $b = 0$, and $b$ is $p–$wave
contribution to $\sigma v$ while $a=0$.

\section{Boltzmann equation for asymmetric dark matter in the quintessence
 model }

In this section we briefly review the relic density computation for asymmetric dark matter in the quintessence model. It was originally appeared in \cite{Iminniyaz:2013cla}. The quintessence model with kination phase has been extensively studied as a class of scalar
field models since the discovery of dark energy \cite{ Salati:2002md, Schelke:2006eg, Catena:2009tm}. It assumes that the reason
behind the current accelerated expansion of the universe is a scalar field
that has minimal coupling to gravity. The ratio of the expansion rate of the
quintessence model with kination phase and the standard model is:
$H^2 /H_{\rm std}^2 = 1 + {\rho}_{\phi} / {\rho}_{r}$, where ${\rho}_{\phi}$
is the scalar energy density. ${\rho}_{\varphi} /\rho_r \approx
k (T / T_{0})^2$ , and $\rho_r = (\pi^2/30)g_*T^4$ is the radiation energy density \cite{ Salati:2002md, Schelke:2006eg}.
Here $T_{0}$ is the reference temperature which is close to
the freezing out temperature,
$k = {\rho}_{\phi} (T_{0} ) /{\rho}_{r} (T_{0})$ . The expansion rate
in the quintessence model can be written as:
\begin{eqnarray} \label{eq:H}
H = A(T) H_{\rm std}\,,
\end{eqnarray}
where $A (T)$ is expressed as $ A(T) = \sqrt{1+ k (T / T_{0})^2}$ \cite{Salati:2002md,Schelke:2006eg}.

Using the modified expression for the expansion rate, we rewrite the Boltzmann
equation (\ref{eq:Yeta}) in quintessence model with kination phase as:
\begin{eqnarray} \label{eq:Q_Yeta}
\frac{d Y^{\pm}}{dx}
    = - \frac{\lambda \langle \sigma v \rangle}{x^2 A(x)}\,
     \left[(Y^{\pm})^2 \mp \eta Y^{\pm} - {Y_{\rm eq}}^2\right]\,.
\end{eqnarray}
We first solve the Boltzmann eqaution (\ref{eq:Q_Yeta}) for $Y^-$, then $Y^+$ is
obtained using the relation $Y^+ - Y^- = \eta$.
For $s-$wave annihilation, the Boltzmann equation for $Y^-$ is
\begin{eqnarray} \label{eq:Qs_Ybareta}
\frac{d Y^-}{dx}
    = - \frac{\lambda a}{x^2 \sqrt{1+ k (x / x_{0})^2}}\,
     \left[(Y^-)^2 + \eta Y^- - {Y_{\rm eq}}^2\right]\,,
\end{eqnarray}

In the case of $p-$wave annihilation, the Boltzmann equation
for $Y^-$ becomes,
\begin{eqnarray} \label{eq:Qp_Ybareta}
\frac{d Y^-}{dx}
    = - \frac{\lambda (6 b / x)}{x^2 \sqrt{1+ k (x / x_{0})^2}}\,
     \left[(Y^-)^2 + \eta Y^- - {Y_{\rm eq}}^2\right]\,.
\end{eqnarray}

We obtain the analytic solution of the relic density for asymmetric dark matter in
quintessence model. We introduce the quantity
$\Delta_- =Y^- - Y_{eq}^-$  . In terms of $\Delta_-$,
Boltzmann equation (\ref{eq:Q_Yeta}) can be rewritten as:
\begin{eqnarray} \label{eq:bigtriangleup_-}
\frac{d \Delta_-}{dx}
    = - \frac{d Y_{eq}^-}{dx} - \frac{\lambda \langle \sigma v \rangle}{x^2 A(x)}
    [\Delta_-( \Delta_- + 2Y_{eq}^-) + \eta \Delta_-]\,.
\end{eqnarray}
For high temperature, the solution for Boltzmann equation (\ref{eq:bigtriangleup_-}) is
\begin{eqnarray} \label{eq:bigtriangleup_-s}
\Delta_- \simeq \frac{2x^2A(x){Y_{eq}}^2}{\lambda \langle \sigma v \rangle ({\eta}^2 +4 {Y_{eq}}^2)}\,.
\end{eqnarray}
Detailed analysis is found in \cite{Iminniyaz:2013cla}. This solution is used to fix the freezing out temperature. When the temperature is low enough, for $x>\bar{x}_F$ ,the term which is proportional to $ Y_{eq}^-$ can be ignored in equation (\ref{eq:bigtriangleup_-}),
so that
\begin{eqnarray} \label{eq:d bigtriangleup_-}
\frac{d \Delta_-}{dx}
     = - \frac{\lambda \langle \sigma v \rangle}{x^2 A(x)}
     (\Delta_-^2 +\eta \Delta_-)\,.
\end{eqnarray}
We may assume $\Delta_- (\bar{x}_F) \gg \Delta_- (\infty)$. Integrating Eq.(\ref{eq:d bigtriangleup_-}) from $\bar{x}_F$ to
$\infty$, then
\begin{eqnarray} \label{eq:Y^-infty}
Y^- (x \to \infty) =
  \frac{\eta}{{\rm exp}[\eta (4\pi / \sqrt{90})m M_{Pl} \sqrt{g_*} I(\bar{x}_F)] - 1}\,.
  \end{eqnarray}
where
\begin{eqnarray} \label{eq:I(x_F^-)}
I(\bar{x}_F) = \int^{\infty}_{\bar{x}_F} \frac{\langle \sigma v \rangle}{x^2 A(x)}dx
=\frac{a}{\sqrt{k}x_0} {\rm ln} \left(\sqrt{k} \frac{x_0}{\bar{x}_F}+ \sqrt{1+ k \frac{x_0^2}{{\bar{x}_F}^2}} \right)
+\frac{6b}{k x_0^2}\left(\sqrt{1+k \frac{x_0^2}{{\bar{x}_F}^2}} -1 \right)\,.
 \end{eqnarray}
Using $\eta =Y^+ -Y^-$, we obtain the relic density of $\chi$ particles.
The result is
\begin{eqnarray} \label{eq:Y^+-infty}
Y^+ (x \to \infty) =
\frac{\eta}{1- {\rm exp}[- \eta (4\pi / \sqrt{90})m M_{Pl} \sqrt{g_*} I(x_F)]}\,.
  \end{eqnarray}
where
\begin{eqnarray} \label{eq:I(x_F)}
I(x_F) = \int^{\infty}_{x_F} \frac{\langle \sigma v \rangle}{x^2 A(x)}dx
=\frac{a}{\sqrt{k}x_0} {\rm ln}\left(\sqrt{k} \frac{x_0}{x_F}+ \sqrt{1+ k \frac{x_0^2}{x_F^2}}\right)
+\frac{6b}{k x_0^2}\left(\sqrt{1+k \frac{x_0^2}{x_F^2}} -1 \right)\,.
 \end{eqnarray}
Eqs.(\ref{eq:Y^-infty}) and (\ref{eq:Y^+-infty}) are consistent with the constraint $\eta =Y^+ -Y^-$ if $x_F = \bar{x}_F$.
We  express the final abundance as
\begin{eqnarray} \label{eq:Omega_DMh^2}
\Omega_{DM}h^2=
\frac{2.76 \times 10^8 m \eta}{{\rm exp}[\eta (4\pi / \sqrt{90})m M_{Pl} \sqrt{g_*}I(\bar{x}_F)]-1} + \frac{2.76 \times 10^8 m \eta}{1-{\rm exp}[-\eta (4\pi / \sqrt{90})m M_{Pl} \sqrt{g_*}I(x_F)]}\,.
\end{eqnarray}
Using the fact that when the deviation $\Delta_-$ is of the same order of the equilibrium value of $Y^-$:
\begin{eqnarray} \label{eq:xi}
\xi Y_{eq}^- (\bar{x}_{F_0}) = \Delta_- (\bar{x}_{F_0})\,.
\end{eqnarray}
The freeze out temperature $\bar{x}_F$ is determined as
\begin{eqnarray} \label{eq:x_F^-}
\bar{x}_F = \bar{x}_{F_0} [1+ \frac{\lambda \eta}{A(\bar{x}_{F_0})} (\frac{0.285a}{(\bar{x}_{F_0})^3} + \frac{1.350b}{(\bar{x}_{F_0})^4})]\,.
\end{eqnarray}
Here $\xi = \sqrt{2} -1$ \cite{Iminniyaz:2013cla}.

\section{Constraints on the parameter spaces}

\begin{figure}[h!]
  \begin{center}
  \hspace*{-0.5cm} \includegraphics*[width=8cm]{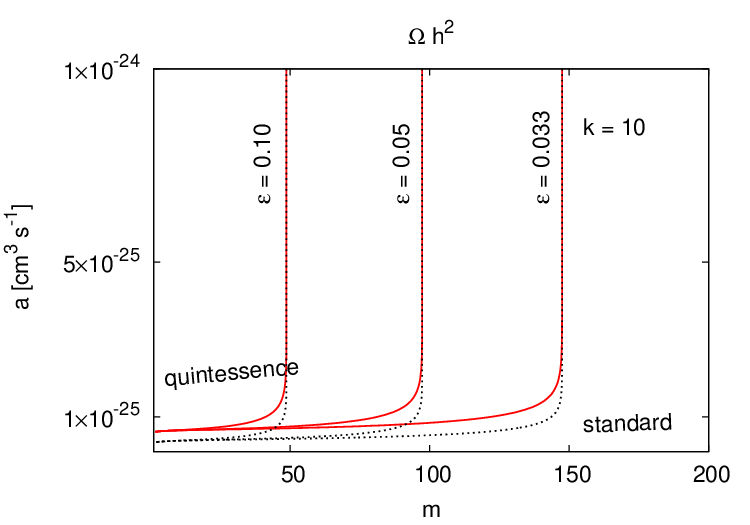}
    \put(-115,-12){(a)}
    \hspace*{-0.5cm} \includegraphics*[width=8cm]{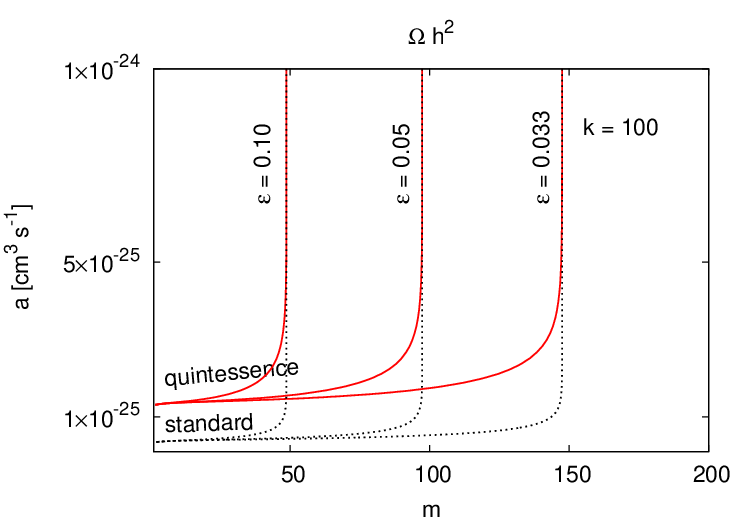}
    \put(-115,-12){(b)}
     \caption{\label{fig:a}
     \footnotesize
  Contour plot of annihilation cross section $a$ and the mass $m$ for different asymmetry when $\Omega_{\rm DM} = 0.120$. Here $\eta  = \epsilon \eta_B $, where $\eta_B = 0.88 \times 10^{-10}$, $b=0$, $g_*=90$, $x_0 = 20$. }
      \end{center}
\end{figure}

\begin{figure}[h!]
  \begin{center}
    \hspace*{-0.5cm} \includegraphics*[width=8cm]{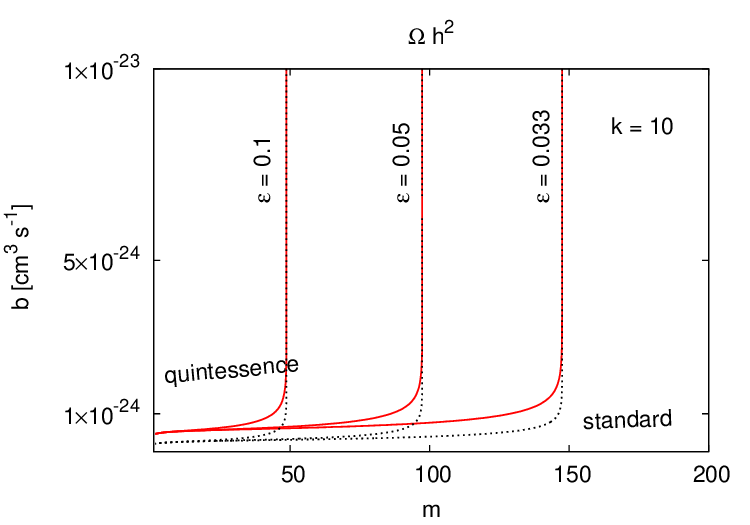}
    \put(-115,-12){(a)}
    \hspace*{-0.5cm} \includegraphics*[width=8cm]{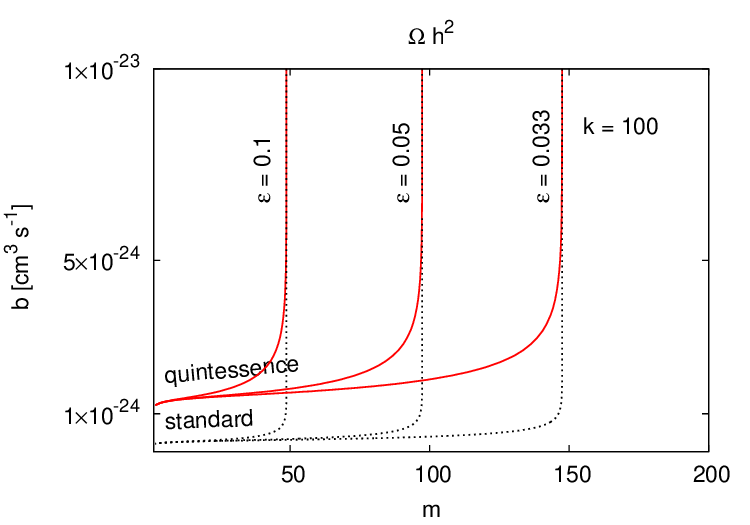}
    \put(-115,-12){(b)}
     \caption{\label{fig:b}
     \footnotesize
  Contour plot of annihilation cross section $b$ and the mass $m$ for different asymmetry when $\Omega_{\rm DM} = 0.120$. Here $\eta  = \epsilon \eta_B $, where $\eta_B = 0.88 \times 10^{-10}$, $a=0$, $g_*=90$, $x_0 = 20$. }
      \end{center}
\end{figure}

Fig.\ref{fig:a} and Fig.\ref{fig:b} are obtained by the numerical solution of Eq.(\ref{eq:Q_Yeta}).
We plot the parameter spaces allowed in $m$--$a(b)$ plane which satisfies the
observed value of dark matter relic density Eq.(\ref{eq:omega}) in the standard cosmological scenario and quintessence model with kination phase for
different parameters $\epsilon$. We take $\epsilon = 0.033\,, 0.05\,, 0.1$. In Fig.\ref{fig:a}, panels $(a)$ and $(b)$ refer to the $s-$wave
processes for $\eta = 10$ and $\eta = 100$ . The solid (red) lines are for the
quintessence model and the dotted (black) lines are for the standard model. Panels $(a)$ and $(b)$ in Fig.\ref{fig:b} are for $p-$wave processes.
We find that for the different asymmetry $\epsilon$, the allowed region is bounded by the cross section $a (b)$ from below and by the maximum value of the mass from the right.
%

For small values of cross sections, the curves are almost flat. When the cross section is small, the symmetric dark matter case is recovered and the mass is irrelevant in that case. This is the reason why the abundance is not sensitive to the varying mass for the small cross section. On the other hand, when the cross section is increased, the abundance is bounded by the
maximum value of the mass from the right. The mass limits are increased when the asymmetry is
decreased. For smaller asymmetry factor the mass bound is larger. For the quintessence model, the lower bound of the cross section is larger than the standard cosmology for the same mass value. The reason for that the Hubble expansion rate is increased in quintessence model. This leads to the larger relic density in the end. Therefore, the cross section should be larger in order to the relic
density falls in the observed value of dark matter abundance.  The same rule
is applied for the case of $p-$wave annihilation. For $p-$wave annihilation,
the cross section limit is one order larger than the $s-$wave annihilation.

\section{Summary and conclusions}
We review the relic density of asymmetric dark matter in the standard
cosmological scenario and quintessence model with kination phase. We use the observed value of dark matter abundance to find
constraints on the parameter spaces including cross section and the mass. We found for
the smaller cross section, the symmetric case is recovered, therefore, the
abundance is not sensitive to the varying mass. When the cross section is
increased, the asymmetric dark matter abundance is decreased and this can be compensated by increasing $m$. The dark matter
abundance is determined almost by the maximum value of the mass and it is not sensitive to the varying cross section. In the case of quintessence model, the cross section is
larger than the standard cosmological scenario due to the increased Hubble
expansion rate. The increased Hubble expansion rate leads to
the larger relic density. The cross section should be
large in order to the dark matter relic density falls in the observed
range. The mass limit is increased for smaller asymmetry factor $\epsilon$.
Our result is important for the asymmetric dark matter models. The
constraints on the parameter spaces like cross section and mass provides the theoretical reference values for the asymmetric dark matter detection experiments.

\section*{Acknowledgments}

The work is supported by the National Natural Science Foundation of China
(2020640017, 11765021) and 2022D01C52.

\end{document}